\documentclass[11pt,a4paper]{article}
\usepackage[utf8]{inputenc}
\usepackage{amsmath}
\usepackage{amsfonts}
\usepackage{amssymb}
\usepackage{graphicx}
\usepackage[per-mode=symbol]{siunitx}
\usepackage{color}
\usepackage{lineno}
\usepackage{caption}
\usepackage{csquotes}
\usepackage{xspace}
\usepackage{eurosym}
\usepackage{booktabs}
\usepackage{hyperref}
\usepackage{cleveref}
\usepackage{authblk}
\DeclareSIUnit{\sieuro}{\mbox{\euro}}
\xspaceaddexceptions{\grqq \grq \csq@qclose@i \} }
\usepackage[textsize=tiny]{todonotes}

\DeclareSIUnit{\bit}{bit}
\DeclareSIUnit{\samples}{sp}
\DeclareSIUnit{\ppm}{ppm}
\author{Philip Hauer\footnote{Corresponding author: hauer@hiskp.uni-bonn.de}, Lars Döpper, Malte Grönbeck,\\Alexander Rachev, Felix Lochter, and Bernhard Ketzer}
\affil{Helmholtz-Institut für Strahlen- und Kernphysik (HISKP)\\Forschungs- und Technologiezentrum Detektorphysik (FTD)\\University of Bonn, Germany}
\DeclareSIUnit{\pixel}{px}
\title{MADHAT: A Sensor Prototype for the ALICE~3 Outer Tracker}

\usepackage[style=numeric,backend=biber,sorting=none]{biblatex} 
\begin{document}
\maketitle
\begin{abstract}
The ALICE collaboration plans a completely new tracking detector based on silicon MAPS technology manufactured in the \SI{65}{\nano\meter} production node as part of the ALICE~3 upgrade during Long Shutdown 4 of the LHC. 
The size of the whole active area is planned to be \SI{50}{\square\meter}, with the Outer Tracker spanning \SI{45}{\square\meter}.

To reach the material budget target of less than \SI{1}{\percent} $X_0$ per layer, open questions like cooling solutions, mechanical structures and electronics have to be answered. 
A simple sensor prototype MADHAT (Mechanical Assessment Design for Heat And Thermal solutions), mechanically identical to the final sensor, but with integrated heating elements and temperature probes, was developed to study the temperature distribution in the final detector assembly and to assess different mechanical designs of the circuit board that houses the sensor.
Furthermore, it will be used to practice the industrialization of the whole mass-production, including the assembly of FPCs with external companies.

In this paper, we present the design of the MADHAT, a prototype readout module and show the results of first measurements.
\end{abstract}

\section{Introduction}\label{Sec:Introduction}
ALICE 3 is a novel, next-generation heavy-ion experiment at the LHC that is planned to be installed during the Long Shutdown 4~\cite{alicecollaborationLetterIntentALICE2022,alicecollaborationScopingDocumentALICE2025}. 
In the new detector, charged-particle tracking will be performed entirely by monolithic active pixel sensors (MAPS) based on \SI{65}{\nano\meter} TPSCo CMOS technology~\cite{HomePageTower2018}.
For the Outer Tracker alone, an area of \SI{45}{\square\meter} needs to be covered, which requires extensive research and development on sensors, mechanics and cooling, as well as cooperation with industry.

The Outer Tracker is split into the barrel and the end-cap regions. In total there will be three barrel layers positioned at a distance of \SI{45}{\cm}, \SI{60}{\cm} and \SI{80}{\cm} from the interaction point, and twelve end-cap disks positioned at a distance of \SI{\pm 77}{\cm}, \SI{\pm 100}{\cm}, \SI{\pm 122}{\cm}, \SI{\pm 150}{\cm}, \SI{\pm 180}{\cm}, and \SI{\pm 220}{\cm} from the interaction point.
Each barrel layer will be \SI{258}{\cm} long, consisting of two staves which meet at the center of the detector at $z=0$. Each stave houses two rows of ten modules, with each module in turn housing eight sensors in a two-by-four configuration.

Especially for the studies about cooling, but also for the cooperation with industry, a placeholder chip is needed.
The placeholder chip should serve two main purposes:
On the one hand, it should be used to emulate a realistic heat dissipation of the final chip so that one can study and evaluate different cooling solutions around it.
A solution based on air cooling is preferred (as it has a very small material budget compared to e.g. a liquid-based solution), but it is not clear whether a stream of air is sufficient to cool the chips and if it introduces vibrations that could deteriorate the performance of the sensors.
In order to measure the temperature of the chips, it should be also equipped with a temperature probe.
On the other hand, it should also be mechanically identical to the final chip so that our partners in industry can train with this chip the different processes, especially the mounting of the chips to modules and the wire-bonding.
More details about the requirements are discussed in \cref{Sec:Requirements}. 

To meet these requirements, we have designed the MADHAT (Mechanical Assessment Design for Heat And Thermal solutions).
It is a simple sensor prototype which is made by applying a single metallization layer onto a silicon wafer.
More details about the design and production of the MADHAT can be found in \cref{Sec:DesignProduction}.
The readout architecture of the module design as well as first measurements will be presented in \cref{Sec:Module}.

\section{Requirements of the MADHAT}\label{Sec:Requirements}
As argued in \cref{Sec:Introduction}, the first requirement of the MADHAT is that it should be mechanically identical to the final chip.
Therefore, the dimensions of the chip are fixed to \SI{25}{\milli\meter} $\times$ \SI{32}{\milli\meter}.
However, the thickness has not yet been decided upon, and the experience we gain in handling MADHATs will help us make this decision.
The normal thickness of a $8''$ or $12''$ wafer is in the order of \SI{750}{\micro\meter}, while the thickness of the epitaxial layer (which forms the active volume of the sensor) is only around \SI{10}{\micro\meter}.
To reduce the amount of inactive material, we plan to thin down the chips to between \SI{50}{\micro\meter} and \SI{100}{\micro\meter}.
Depending on the thickness, the handling of chips becomes different.
For example, silicon becomes flexible when the thickness of a chip is significantly less than \SI{100}{\micro\meter}.
Therefore, we decided to thin down some MADHATs to \SI{60}{\micro\meter} and some to \SI{100}{\micro\meter}.
For testing the reliable electric connection between module PCB and MADHAT, additional pads for wire-bonding are necessary.

The second requirement of the MADHAT is the emulation of a realistic heat profile to study different cooling solutions.
Furthermore, we want to implement a temperature probe on the MADHAT itself.
This way, we can directly measure the temperature on the chip surface and do not have to rely on external methods (e.g. an infrared camera) or have to solder additional components onto the chip (e.g. a PT100).
For both -- heating structure and temperature probe -- we want to use a single metallization layer on the chip made of aluminium.
The metal layer is then patterned in a photo-lithographic process to form meandering resistive structures.
By driving a current through the heating lines, power is dissipated on the chip through the finite conductivity of the heating lines.
When the chip temperature changes, the resistivity of the metal layer also changes, which we use for the temperature probe by measuring the resistance of the metal lines used for the temperature probes.
More details will be explained in \cref{Sec:DesignProduction}.

In order to create a realistic heat profile, we distinguish between the periphery area and the matrix area.
The periphery is the part of the chip where the connection to the module and signal processing takes place.
It is located on the long side of the chip and we expect it to be \SI{1.5}{\milli\meter} wide.
Since all signals and power lines are routed via the periphery, we expect the power density to be \SI{135}{\milli\watt\per\square\centi\meter}.
The pixels of the chip are the building blocks of the matrix area, their power density is expected to be around \SI{30}{\milli\watt\per\square\centi\meter}.
This was taken into account during the design of the MADHAT, i.e. the heating structure is more dense in the periphery area than in the matrix area.

\section{Design and Production of the MADHAT}\label{Sec:DesignProduction}
\subsection{Design of the MADHAT}\label{Sec:DesignProduction:Design}
\begin{figure}
    \centering
    \includegraphics[width=\textwidth]{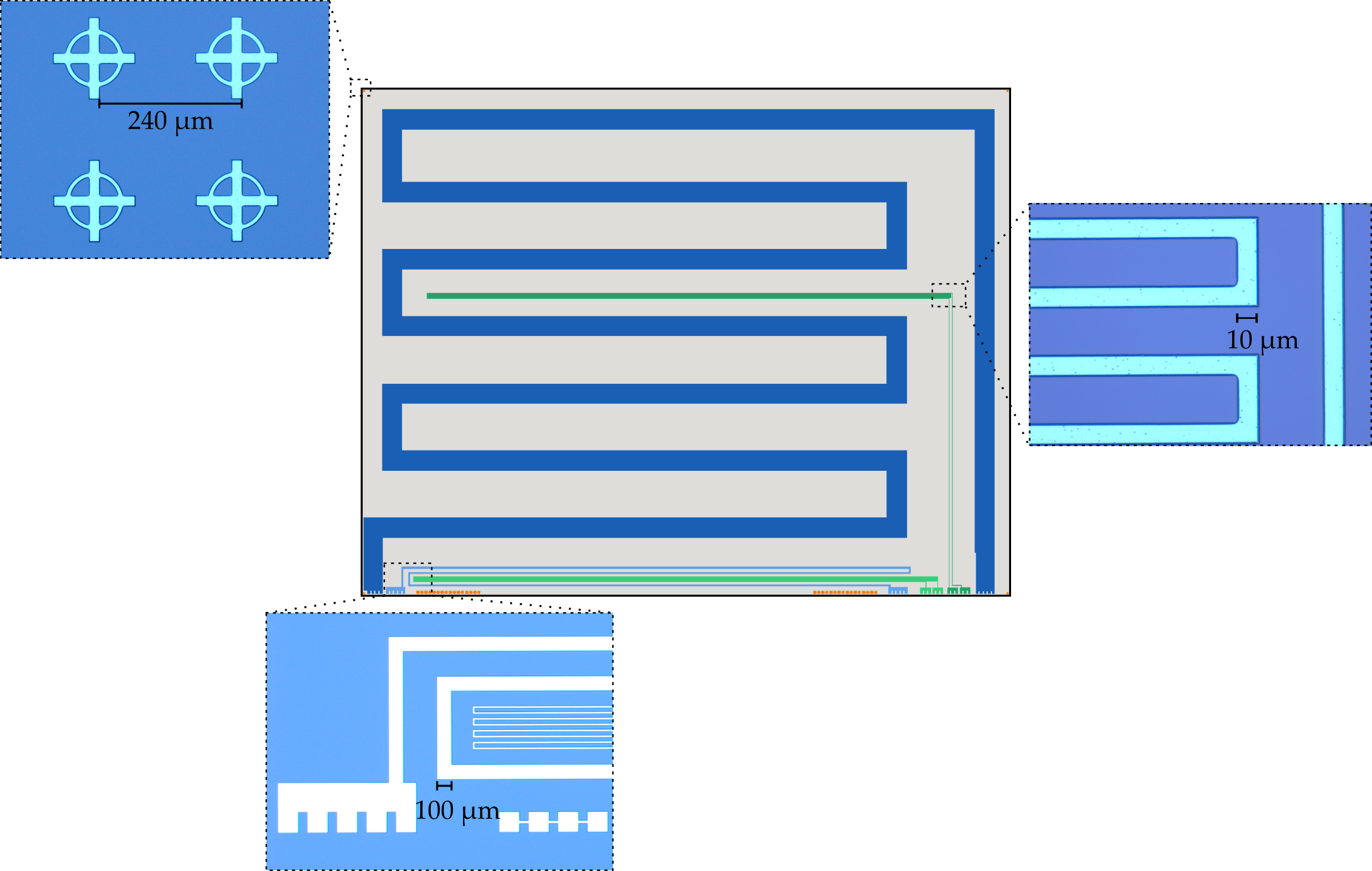}
    \caption{The MADHAT (sketch) with three close-up views (photos).
    The thick meandering structures are for the heat dissipation (dark blue for the matrix and light blue for the periphery), the thin meandering lines are for the measurement of the temperature (dark green for the matrix, light green for the periphery).
    Furthermore, there are two passive structures (depicted in orange): Additional bonding pads in the periphery region and alignment markers in the four corners of the chip.\\
    The three close-up views are microscopic pictures of the respective area.
    The top left one shows the alignment markers on four neighbouring chips, the right one shows the meandering structure of the thermal probe line of the matrix region and the one on the bottom shows a part of the heating structure and temperature probe for the periphery region as well as additional pads for wire-bonding.}
    \label{Fig:Design:MADHATSketch}
\end{figure}

As discussed in \cref{Sec:Requirements}, MADHATs are produced with silicon wafers with one metallization layer.
The silicon wafer has a \SI{500}{\nano\meter} thick SiO$_2$ layer to prevent a direct connection between silicon and metal.
The design of the metallization pattern is shown in \cref{Fig:Design:MADHATSketch}.

For the design, we distinguish between the matrix and the periphery region (see \cref{Sec:Requirements}).
For both regions, a realistic heat profile should be emulated.
Therefore, two thicker lines of aluminium -- one for the matrix and one for the periphery -- are applied onto the chip.
Since the power density is higher in the periphery, the lines are thinner such that more power is dissipated there.

\begin{table}[]
    \centering
    \begin{tabular}{c|c|c|c|c}
         Trace & Width  & Length & Resistance  & Dissipated power \\
         \midrule
         Matrix power & $\SI{1000}{\um}$ & $\SI{220}{\mm}$ & $\SI{6.4}{\ohm}$ & $\SI{65.8}{\milli\watt}$ \\
         Periphery power & $\SI{100}{\um}$ & $\SI{75}{\mm}$ & $\SI{21.9}{\ohm}$ & $\SI{22.6}{\milli\watt}$ \\
         Matrix probe & $\SI{15}{\um}$ & $\SI{236}{\mm}$ & $\SI{457}{\ohm}$ & $\SI{4.6}{\micro\watt}$\\
         Periphery probe & $\SI{15}{\um}$ & $\SI{207}{\mm}$ & $\SI{400}{\ohm}$ & $\SI{4}{\micro\watt}$\\
    \end{tabular}
    \caption{Target values of aluminium traces. The dissipated power is calculated for the target voltage of $\SI{1.2}{\V}$ (for both power regions) and $\SI{100}{\micro\A}$ of constant current (for both probe regions).}
    \label{tab:targetValues}
\end{table}
In order to measure the temperature of the chip, very thin and long traces of aluminium are patterned onto the chip.
When driving a constant current through these lines, a voltage drop, which is proportional to the resistance, occurs.
As the resistance depends on the temperature, the voltage drop is a measure for the temperature of the chip.
All aluminium trace target values are shown in \cref{tab:targetValues}.


\subsection{Production of the MADHAT}\label{Sec:DesignProduction:Production}
\begin{figure}
\begin{minipage}[t]{0.47\textwidth}
    \centering
    \includegraphics[width=\linewidth]{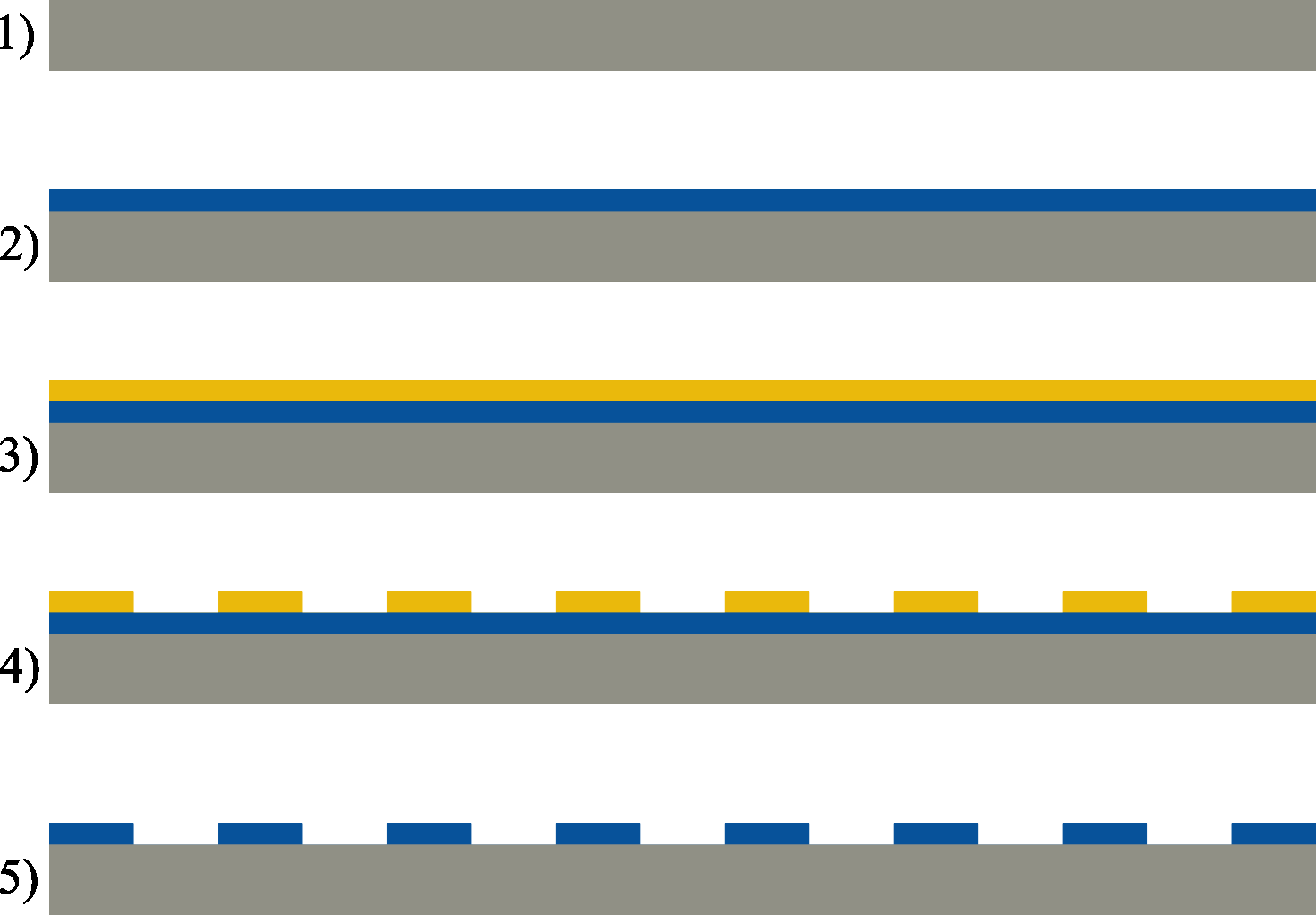}
    \caption{Schematic of the production process.
    Explanation in text.}
    \label{Fig:Design:ProductionSchematic}
\end{minipage}\hfill
\begin{minipage}[t]{0.47\textwidth}
    \centering
    \includegraphics[width=\linewidth]{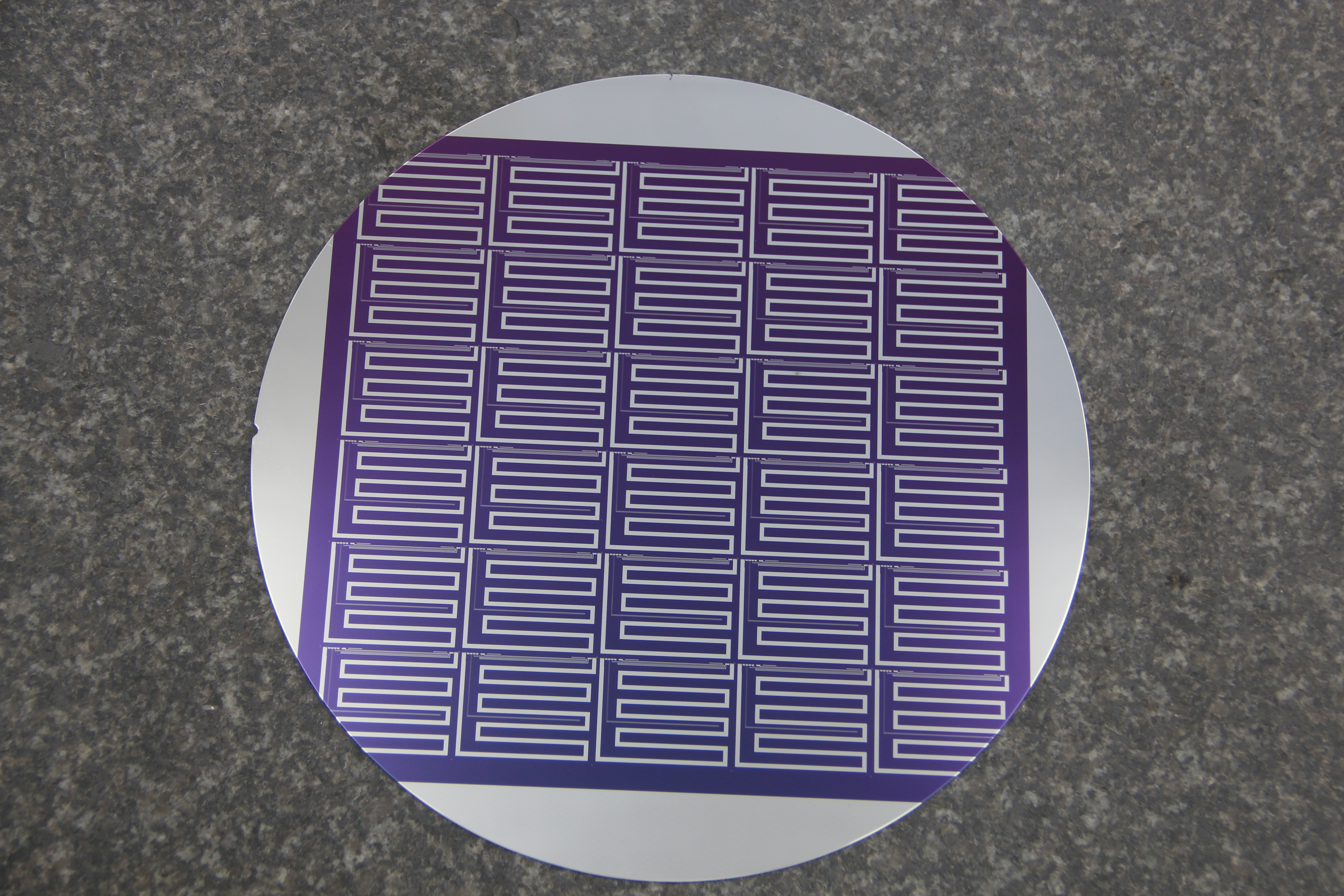}
    \caption{A \SI{200}{\milli\meter} wafer with 26 MADHATs.}
    \label{Fig:Design:ProductionFullWafer}
\end{minipage}
\end{figure}

The production of the first MADHATs took place at the Forschungs- und Technologiezentrum Detektorphysik (FTD) of the University of Bonn.
The FTD provides a large cleanroom facility with a dedicated area for photolithography.
This area houses different machines that were used for the production of the MADHATs.

A schematic of the process is depicted in \cref{Fig:Design:ProductionSchematic}.
The base material (grey area in 1) is a silicon wafer with a diameter of \SI{200}{\milli\meter} and with a \SI{500}{\nm} thick SiO$_2$ layer on top.
In a first step, a \SI{1000}{\nano\meter} thick layer of aluminium is sputtered onto the wafer (see blue area in 2).
Afterwards, the fully metalized wafer is mounted into a spin-coater in order to apply the liquid photoresist AZ1505 by MicroChemicals~\cite{merckkgaaTechnicalDatasheetAZ2021} with a thickness of $\approx \! \SI{1}{\um}$ (yellow area in 3).
Next, the photoresist has to be patterned (holes in the yellow area in 4).
For this, we are using the MLA150 by the company Heidelberg Instruments~\cite{heidelberginstrumentsMLA150Advanced2024}.
It uses a UV laser to directly pattern the photoresist without the need of a mask.
In the next step, the exposed aluminium is etched away with an acid based on phosphoric acid.
Lastly, the photoresist is stripped away with acetone (5).

Each wafer can hold 26 MADHATs as depicted in \cref{Fig:Design:ProductionFullWafer}.
In total, we produced 25 wafers.
In order to further process these wafers, they have to be thinned and diced.
For first tests, we took two wafers and did not thin them down but only diced them with the wafer saw from type DAD3660~\cite{discocorporationDicingSawDAD3660} in the FTD.
Results of these sensors will be discussed in \cref{Sec:Module:FirstPrototype}.
As the FTD has no capabilities to thin down wafers, we contacted external companies to perform this service for us.
For a first test, we sent two wafers to the company DISCO~\cite{DISCOCORPORATION}, which thinned down one wafer to a thickness of \SI{60}{\micro\meter} and one wafer to a thickness of \SI{100}{\micro\meter} and diced both of them.
They performed the thinning and dicing process with the so-called \enquote{dicing before grinding} process, where in a first step grooves are cut into the wafer and in the second step the thinning is performed, separating the chips at the grove lines.

After inspecting the quality of the chips received from DISCO, we also sent the remaining 21 wafers for thinning and dicing to them.
14 of them were thinned down to \SI{60}{\micro\meter}, seven of them to \SI{100}{\micro\meter}.
Further tests about cooling and vibration studies will be conducted soon, but are not part of this paper.


\section{Module Design}\label{Sec:Module}

\subsection{Readout circuit}\label{Sec:Module:ReadoutCircuit}
An important requirement for the MADHAT is the possibility not only to create heat but also to measure the temperature of the chip.
For this, each MADHAT houses two thin aluminium traces.
These lines act as temperature-dependent resistors.
One resistor is for reading the temperature of the matrix ($R_\textrm{matrix}$) and one resistor for the periphery ($R_\textrm{periphery}$).

\begin{figure}
    \centering
    \includegraphics[width=0.8\linewidth]{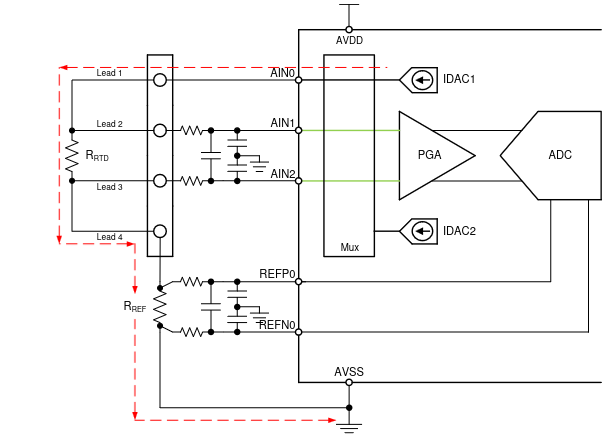}
    \caption{Schematic of four-wire sensing. The red dashed line depicts the path of the constant current source. Taken from \cite{wuBasicGuideRTD2024}.}
    \label{fig:4wire}
\end{figure}

\begin{figure}
\begin{minipage}[t]{\textwidth}
    \centering
    \includegraphics[width=\linewidth]{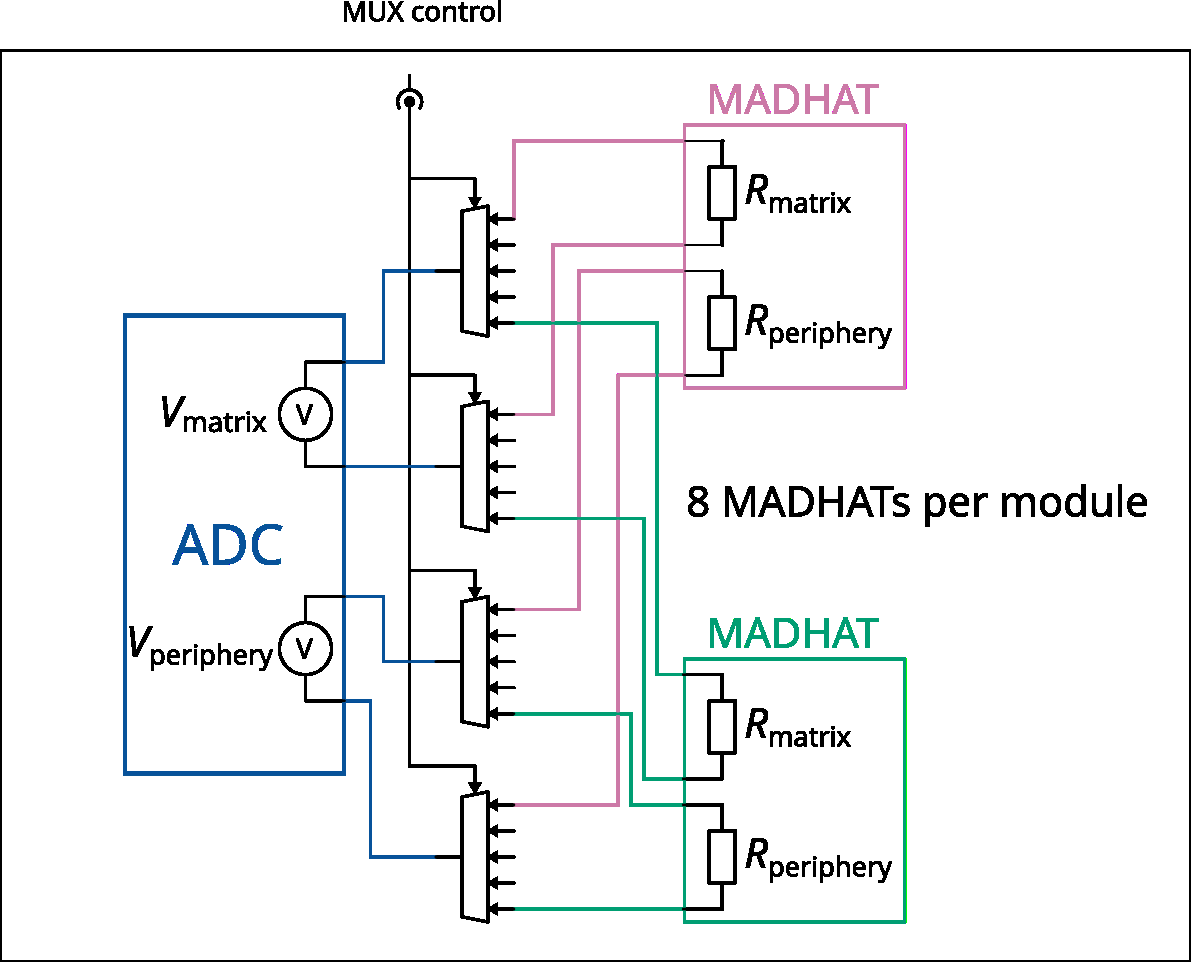}
    \caption{Schematic design of the readout architecture for the temperature probes.
    Explanation in text.}
    \label{Fig:Module:ReadoutCircuit}
\end{minipage}
\end{figure}

In order to measure the temperature, one therefore has to measure the resistance of the two resistive lines and compare it to calibrated values.
A common way to perform these measurements is the so-called four-wire sensing, which is sketched in \cref{fig:4wire}.
By using separate pairs of current-carrying and voltage-sensing electrodes, one can mitigate the influence of the connections to the MADHAT.
Another important aspect of the module design is the minimization of electric components on the backside as this would influence the measurements for the cooling studies (especially the measurements of vibrations).
To read out the 8 MADHATs (16 probes) per module, we decided to use four multiplexers with eight channels and one 16-bit ADC with four channels\footnote{The ADC we selected has six channels in total, of which four are used to measure the voltage drop over the temperature probes, the other two channels are used to supply a constant current to the temperature probes.} as depicted in \cref{Fig:Module:ReadoutCircuit}.
The multiplexers can be controlled by the bus line (indicated with \enquote{MUX control}) and thus the number of components on the module is minimized.

\subsection{First Prototype}\label{Sec:Module:FirstPrototype}
\begin{figure}
\hfill
\begin{minipage}[t]{0.45\textwidth}
    \centering
    \includegraphics[width=\linewidth]{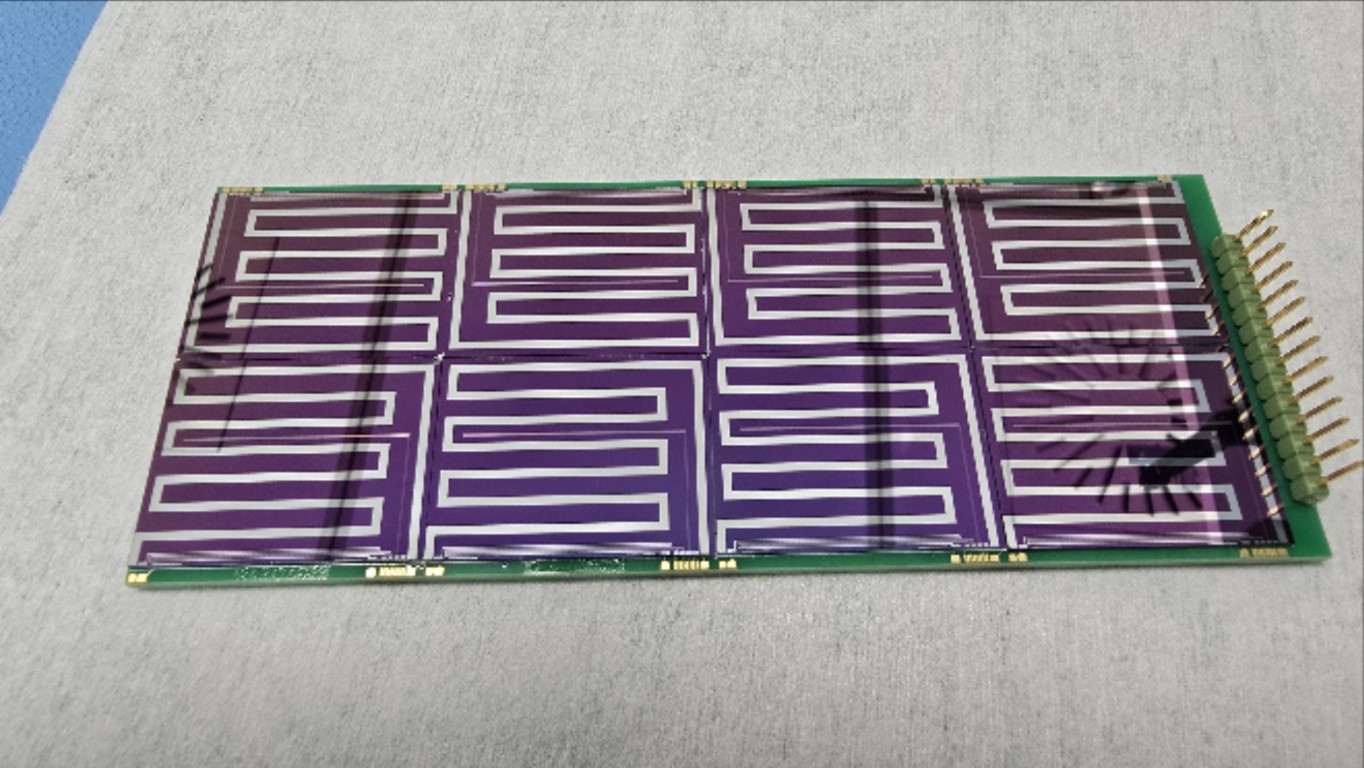}
    \caption{Eight MADHATs glued and wire-bonded onto a module with a pin-header to connect to the ESP32 for power and data transmission.}
    \label{Fig:Module:FirstPrototype}
\end{minipage}\hspace{0.05\linewidth}
\begin{minipage}[t]{0.45\textwidth}
    \centering
    \includegraphics[width=\linewidth]{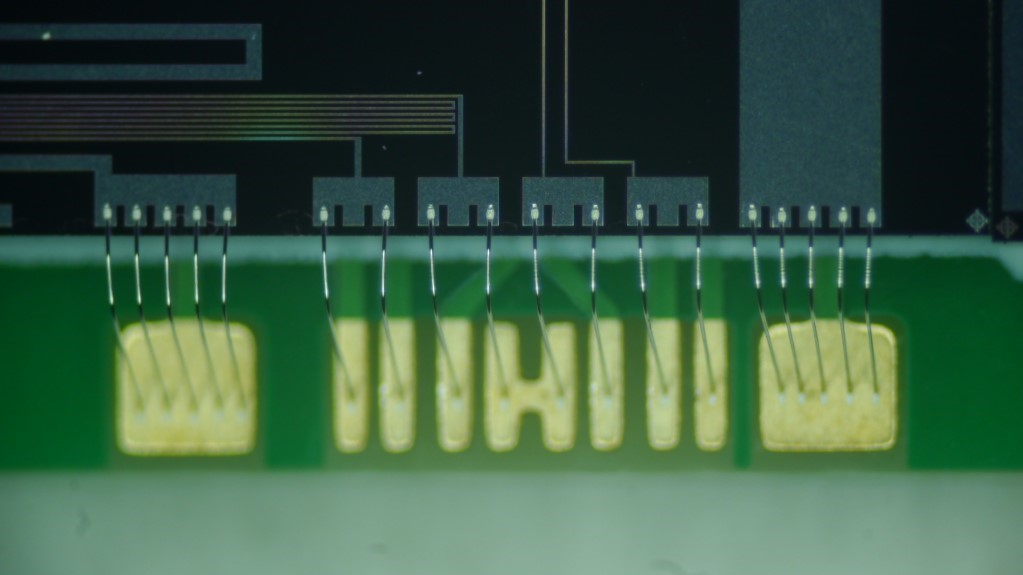}
    \caption{Wire-bonded connection between MADHAT and the module.
    Also visible: The heating structures (thick lines) and temperature probes (thin lines).}
    \label{Fig:Module:WireBonds}
\end{minipage}
\hfill
\end{figure}

A first prototype module has been assembled by glueing eight MADHATs onto a rigid PCB.
The setup is depicted in \cref{Fig:Module:FirstPrototype}.
The connection between MADHAT and module has been established by wire-bonds between chip and module.
A close-up view of the wire-bond connection is shown in \cref{Fig:Module:WireBonds}.

Not shown in the picture is the microcontroller of type ESP32, which will be connected via standard jumper cables to the pin header on the module.
With it, the power for the heating structure is generated as well as the data transmission from and to the MADHATs. For this first prototype only one module can be connected and read out. 

\subsection{Temperature Probe Calibration}
\begin{figure}
\centering
\includegraphics[width=0.6\linewidth]{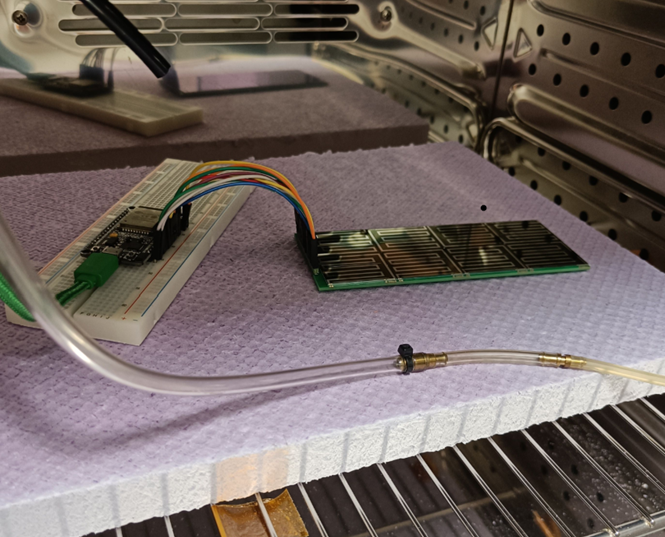}
\caption{The first MADHAT prototype in a climate chamber for calibration.}
\label{fig:Prototype:ClimateChamber}
\end{figure}

A first step in the commissioning of the prototype is the calibration of the temperature probes with an external device.
For this, we have chosen the climate chamber Binder MK 56~\cite{binderBinderKB65} to scan the temperature in the range from \SIrange{5}{45}{\degreeCelsius} in steps of \SI{2}{\degreeCelsius}.
The setup is depicted in \cref{fig:Prototype:ClimateChamber}.
At each temperature step, after heating up or cooling down, the temperature was kept constant for \SI{20}{\minute} to reach the thermal equilibrium.
Afterwards, the voltage drop across the temperature probes of the MADHAT has been measured by the ADC.
Note that the heating structures of the MADHAT have not been used for this calibration.
In total, the whole calibration process took around \SI{20}{\hour}. This time can be reduced to just \SIrange{2}{3}{\hour}, as such short steps are not necessary for the needed precision.

\begin{figure}
\centering
\includegraphics[width=\linewidth]{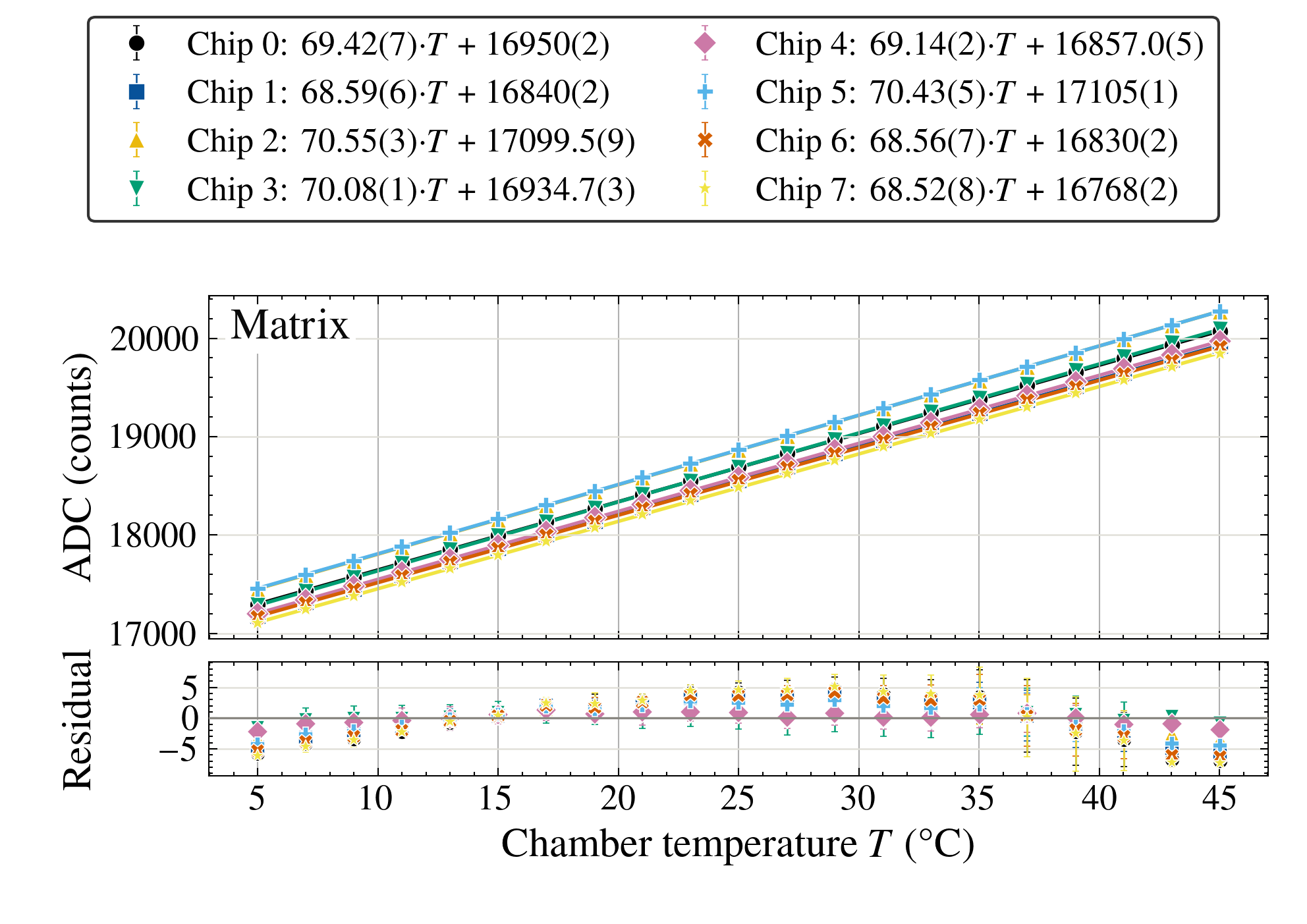}
\caption{Calibration results for all eight MADHATs (matrix regions) on the prototype module.}
\label{fig:Prototype:CalibrationResults}
\end{figure}

The measurement results for the measurements probes of the eight matrix regions are exemplarily depicted in \cref{fig:Prototype:CalibrationResults}.
On the x-axis, the temperature of the climate chamber is shown, while the measured ADC values for the matrix region are displayed on the y-axis for every chip.
A linear dependency between the measured ADC values of each chip and the temperature of the climate chamber is visible as indicated by the fit.
Even though the residuals hint that a quadratic fit might be better suited, the precision of the temperature measurement is better than \SI{0.1}{\degreeCelsius} (which can be estimated by the largest slope and the largest residual).
However, each chip has slightly different resistances which might originate from imperfections in the production process, possibly inhomogeneity of the sputtering.
Therefore a prior calibration of each MADHAT and individual conversion parameters to accurately measure temperature are required.

\subsection{Verification of the Heating Structure}
\begin{figure}
    \centering
    \includegraphics[width=0.5\linewidth]{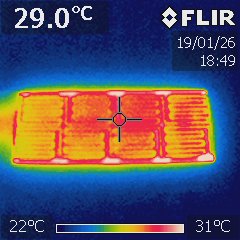}
    \caption{Image of the prototype module with a thermal-imaging camera.}
    \label{fig:Thermal_Image}
\end{figure}

To verify the functionality of the heating structures, all eight MADHATs of the prototype module were powered with normal operating conditions and observed with a thermal-imaging camera. 
The resulting infrared image is shown in \cref{fig:Thermal_Image}. 
All eight chips heat up uniformly and are clearly distinguishable from the cooler carrier PCB, reaching surface temperatures of up to \SI{31}{\degreeCelsius} at an ambient temperature of around \SI{20}{\degreeCelsius}. 
The meandering heating lines of the matrix region are individually resolved as a striped pattern on each chip, and the higher power density of the periphery region is visible as a distinctly warmer band along the long side of each chip. 
This confirms that the heating structures produce the intended two-zone heat profile described in section 2. 

\section{Summary and Outlook}\label{Sec:SummaryOutlook}
In this manuscript, we presented the development and initial validation of the MADHAT -- the mechanic and thermic equivalent for the ALICE~3 Outer Tracker chip.
It will serve two main purposes: 
First, it will be used to study cooling solutions and second, it will be used to qualify industrial assembly processes.
The MADHAT design, implemented using a single metallization layer on a silicon wafer, integrates heating and precision temperature-sensing structures.
With the machines and facilities provided by the FTD, we were able to produce the chips in-house.

We assembled a first prototype module with eight MADHAT chips.
The setup allowed us to characterize the embedded temperature probes, confirming a precision of better than \SI{0.1}{\degreeCelsius}. 
The overall accuracy of the temperature probes is limited by the accuracy of the climate chamber, and can be estimated to be around \SI{1}{\degreeCelsius}.
The performance of the heating structure was also verified.
However, a meaningful analysis of heat dissipation requires a more complete system, as the thermal performance depends on the interplay of several modules within the stave structure.

Our immediate focus is on constructing complete staves with thinned MADHAT chips to perform systematic tests of various cooling solutions.
Particular emphasis will be placed on evaluating air cooling solutions, as they typically have the smallest material budget.
A key aspect of this study will be to characterize and mitigate potential vibrations induced by the coolant flow.
For these measurements, it is also indispensable to ensure the realistic thickness of the chips.
The thinning of the wafers can not be done at the FTD and we have to rely on industrial partners.
For the next iteration, we want to use flexible printed circuit boards instead of the rigid ones that we used for the first prototype module.
Again, this is also important for the next measurements that we will perform with these modules.

Concurrently, the MADHAT will serve as a vital tool for qualifying industrial production chains.
We will partner with companies to fabricate the chips and assemble them onto flexible printed circuit boards, including the wire-bonding.
The MADHAT is ideally suited for this case:
It is a low-cost yet mechanically and thermally realistic substitute for the final chip, allowing for thorough process testing without the expense of prototype chips.

The successful execution of this program will de-risk critical aspects of the ALICE~3 Outer Tracker construction, specifically thermal management and large-scale industrial assembly, thereby providing essential input for the final detector design.

\section*{Acknowledgments}
This work was supported by the German Federal Ministry for Research, Technology and Space (BMFTR) under grant number 05P24PD3.
The Research and Technology Center for Detector Physics (FTD) at the University of Bonn was jointly funded by the federal government and the state of North Rhine-Westphalia as part of the program for research facilities under Article 91b of the German Grundgesetz.


\printbibliography

@misc{alicecollaborationLetterIntentALICE2022,
  title = {Letter of Intent for {{ALICE}} 3: {{A}} next-Generation Heavy-Ion Experiment at the {{LHC}}},
  author = {{ALICE Collaboration}},
  date = {2022},
  location = {Geneva},
  url = {https://cds.cern.ch/record/2803563}
}

@misc{alicecollaborationScopingDocumentALICE2025,
  title = {Scoping Document for the {{ALICE}} 3 Detector},
  author = {{ALICE Collaboration}},
  date = {2025},
  location = {Geneva},
  url = {https://cds.cern.ch/record/2925455},
  organization = {CERN}
}

@online{binderBinderKB65,
  title = {Binder MK 56},
  author = {{Binder}},
  url = {https://www.binder-world.com/de-de/produkte/klimaschraenke/wechselklimaschraenke/produkt/mk-56},
  urldate = {2026-04-28},
  langid = {german}
}

@online{DISCOCORPORATION,
  title = {{{DISCO CORPORATION}}},
  url = {https://www.disco.co.jp/eg/index.html},
  urldate = {2026-04-28}
}

@online{discocorporationDicingSawDAD3660,
  title = {Dicing {{Saw DAD3660}}},
  author = {{DISCO Corporation}},
  url = {https://www.disco.co.jp/eg/products/dicer/dad3660.html},
  urldate = {2026-04-28}
}

@misc{heidelberginstrumentsMLA150Advanced2024,
  title = {{{MLA}} 150 - {{The}} Advanced Maskless Aligner},
  author = {{Heidelberg Instruments}},
  date = {2024-06-14},
  url = {https://heidelberg-instruments.com/product/mla150/},
  urldate = {2024-06-14},
  langid = {british}
}

@online{HomePageTower2018,
  title = {Tower {{Partners Semiconductor Co}}.},
  date = {2018-02-08},
  url = {https://towersemi.com/},
  urldate = {2026-04-28},
  langid = {american}
}

@misc{merckkgaaTechnicalDatasheetAZ2021,
  title = {Technical {{Datasheet}} - {{AZ}} 1500 Series},
  author = {{Merck KGaA}},
  date = {2021},
  url = {https://www.microchemicals.com/dokumente/datenblaetter/tds/merck/en/tds_az_1500_series.pdf},
  urldate = {2024-09-19},
  organization = {Merck KGaA}
}

@misc{wuBasicGuideRTD2024,
  title = {A {{Basic Guide}} to {{RTD Measurements}}},
  author = {Wu, Joseph},
  date = {2026-04-28},
  url = {https://www.ti.com/document-viewer/lit/html/sbaa275},
  urldate = {2026-04-28},
  organization = {Texas Instruments}
}
  
\end{document}